\documentstyle[11pt,a4]{article}
\begin{document}
%\section{Introduction}

\indent

In order to realize the main physics goal at the future
hadronic colliders LHC and SSC,
the precise estimate of inclusive production of neutral clusters
is needed to pin down signals due to Higgs particle or new
physics \cite{altap}.
For this purpose leading order (hereafter denoted as LO )
perturbative QCD
predictions - based on
evaluations of partonic cross sections at tree level and evolution of
structure and fragmentation functions at one loop level--~
are not accurate enough.
A consistent calculation at next to leading order
(hereafter denoted as NLO) needs two loop evolved structure and fragmentation
functions and a NLO evaluation of
parton-parton subprocesses which has been performed by Aversa et al.
\cite{acgg}  a few years
ago and is based on the O($\alpha_s^3$)
matrix elements of Ellis and Sexton\cite{elliss}.

\indent

Up to now such an analysis has been performed for
for heavy quarks \cite{melna} and pion
fragmentation functions \cite{ftn}.
Following the latter approach we present
a complete NLO evaluation of eta inclusive
production at hadronic colliders in order to estimate, as precisely
as possible, the
$\eta$ rates at LHC  and SSC.

\indent
We first extract $\eta$ fragmentation functions from $e^+e^-$ and $p\bar p$
collisions  at the scale $M_{f_0}=$30 GeV
using one of the three methods developed in our previous
work\cite{ftn}, where we have shown that those different approaches give
consistent theoretical predictions
for $\pi^0$ production
at LHC/SSC. Therefore we are quite
confident on the accuracy of the procedure followed below
for $\eta$.

Then we will confront our predictions with the existing experimental
data in $e^+e^-$ and $p\bar p$ annihilation in order to
estimate the theoretical uncertainty.
Cross-sections for $\eta$ production in hadron collisions
are usually not directly measured,
but they
are deduced from $\pi^0$ cross-sections, assuming an experimental
$\eta/\pi^0$ ratio which has been measured at  ISR\cite{afs}
and is taken to be constant of order 0.5.

We find that the theoretical $\eta/\pi^0$ ratio increases with
$p_t$ at present energies, and therefore  we
study in detail this ratio in the LHC/SSC energy range, when we give our
predictions for the future colliders.

\indent

We briefly report, for reader's convenience, the main formulae for one hadron
inclusive
production at
next-to-leading-order\cite{ftn}, via the generic reaction $ A+B\to H
+ X$,
where A
and B stand for hadrons and/or leptons. The cross section is given by
the convolution of the partonic cross section and the fragmentation
functions $D_l^H(z,M_f^2)$:
\begin{equation}
E_{H} \; \frac{d\sigma_{A+B \rightarrow H}}{d^3 \vec{P}_{H}}
 =  \sum_{l} \; \int_{z_H}^{1} \frac{dz}{z^2} \; D^{H}_{l}(z,
M^2_f) \; E_l \: \frac{d\sigma_{A+B \rightarrow l}}{d^3 \vec{P}_{l}}
(\frac{z_H}{z}, \theta, \alpha_s(\mu^2), M^2_f,\cdots) ,
\label{meq}
\end{equation}
where $z_H$ is the reduced energy of the hadron $H$,
$z_H=2 E_H / \sqrt S$,
$\theta$ is the scattered angle of the parton l, and
the inclusive production of the parton l in the reaction
$A+B \rightarrow l$
has the following perturbative development:
\begin{equation}
E_l \; \frac{d\sigma_{A+B \rightarrow l}}{d^3 \vec{P}_{l}}
(\frac{z_H}{z}, \theta, \alpha_s(\mu^2), M^2_f,\cdots ) =
\sigma^0_{A+B \rightarrow l}(\frac{z_H}{z},\theta) + \frac{\alpha_s
(\mu^2)}
{2 \pi} \; \sigma^1_{A+B \rightarrow l}(\frac{z_H}{z},\theta,M^2_f) +
\cdots .
\end{equation}
Furthermore $D^H_l(z,M^2_f)$ represents the number of hadrons H
inside the parton l carrying the fraction of momentum $z$ from H,
evolved at the scale $M^2_f$.
These fragmentation functions satisfy Altarelli-Parisi type
evolution equations \cite{altap}.

\indent

For inclusive production in $e^+e^-$ annihilation, the
partonic cross-sections at next-to-leading order are given by:
\begin{eqnarray}
\lefteqn{E_{q_i} \: \frac{d\sigma_{e^+ + e^- \rightarrow q_i}}{d^3
\vec{P}_{q_i}} (y,\theta,\alpha_s(\mu^2),M^2_f) = } \nonumber \\
&  & \frac{6 \: \sigma_0}{\pi Q^2 y}
\; e_i^2 \left\{ \frac{3}{8} (1+\cos^2 \theta) \left[
\delta(1-y) + \frac{\alpha_s(\mu^2)}{2 \pi} \left(
P^0_{qq}(y) \ln\!\left( \frac{Q^2}{M^2_f} \right) + K_q^T(y) \right)
\right] \right. \nonumber \\
& & \mbox{} + \left. \frac{3}{4} (1-\cos^2 \theta)
\frac{\alpha_s(\mu^2)}{2 \pi} K^L_q(y) \right\} \\
\lefteqn{E_g \: \frac{d\sigma_{e^+ + e^- \rightarrow g}}{d^3 \vec{P}_{g}}
(y,\theta,\alpha_s(\mu^2),M_f^2) = } \nonumber \\
&  & \frac{12 \: \sigma_0}{\pi Q^2 y}
\sum_{i=u,d,s,c,...} e_i^2
\left\{ \frac{3}{8} (1+\cos^2 \theta)
\left[ \frac{\alpha_s(\mu^2)}{2 \pi} \left(
P^0_{gq}(y) \ln\!\left( \frac{Q^2}{M^2_f} \right) + K_g^T(y) \right)
\right] \right. \nonumber \\
& & \mbox{} + \left. \frac{3}{4} (1-\cos^2 \theta)
\frac{\alpha_s(\mu^2)}{2 \pi} K^L_g(y) \right\} ,
\end{eqnarray}
where $\sigma_0$ is the usual point like cross-section
\[ \sigma_0 = \frac{4 \pi \alpha^2}{3 Q^2} , \]
$\alpha$ is the QED coupling constant and $Q^2$ is the invariant
mass of the $e^+ e^-$ pair.
The functions $K^T_q$, $K^L_q$, $K^T_g$ and $K^L_g$ have
been extracted from the
reference \cite{alta1} (see also \cite{others}) and are given
elsewhere\cite{ftn}.

\indent

For hadronic collisions the partonic cross-sections are:
\begin{eqnarray}
E_l \frac{d\sigma_{p+p \rightarrow l}}{d^3 \vec{P}_{l}}
(y,\theta,\alpha_s(\mu^2),M_f^2) & = &
\frac{1}{\pi S} \sum_{i,j} \int_{V W}^{V} \frac{dv}{1-v}
\int_{V W/v}^{1} \frac{dw}{w} \nonumber \\
& & \mbox{} \times \left[ F^p_i(x_1,M^2)
F^p_j(x_2,M^2) \left( \frac{1}{v} \left( \frac{d \sigma^0}{dv}
\right)_{i j \rightarrow l} (s,v) \delta(1-w) \right. \right. \nonumber
\\
& & \left. \left. + \frac{\alpha_s(\mu^2)}{2 \pi} K_{i j \rightarrow l}
(s,v,w;\mu^2;M^2,M_f^2) \right) + (x_1 \leftrightarrow x_2) \right] ,
\end{eqnarray}
where the partonic variables are $s=x_1x_2S$ and
\[ x_1= \frac{VW}{vw} , \; \; x_2  = \frac{1-V}{1-v}, \]
and the hadronic ones are defined by:
\[ V = 1- \frac{y}{2} (1-\cos \theta), \; \; W= \frac{y (1+\cos \theta)}
{2 - y (1-\cos \theta)} . \]

\indent

We now consider the $\eta$ inclusive production in $e^+e^-$
annihilation at $M_{f0}=\sqrt S=30$ GeV, as simulated by the Monte Carlo
generator HERWIG\cite{her}. As well known, this event generator includes the
QCD
parton shower to leading and next to leading accuracy - in particular
the kinematical corrections due to the phase space boundaries are
summed up to all orders - as well as the
hadronisation of the color singlet clusters into the physical particles.
Furthermore HERWIG has been shown \cite{herlep} to describe with good
accuracy the observed features of PETRA and LEP data.
Then, similarly to what have been done for $\pi^0$ inclusive production,
we will use the $\eta$ distribution
generated by each quark flavor which originates from the
photonic vertex, as a
realistic description of the quark fragmentation into $\eta$.
Owing to the
symmetry of quarks and antiquarks fragmenting into $\eta$
we extract the quark fragmentation functions from:
\begin{equation}
{d\sigma_{e^+e^-} \rightarrow \eta \over dz_H} (z_H,M_{f0}^2) \sim
 6 \sigma_0 \sum_q e_q^2 D_q^{\eta} (z_H, M_{f0}^2),
\label{her1}
\end{equation}
where $\sigma_0$ is the pointlike cross section defined above.

The reaction $ e^+e^- \rightarrow \eta + X $ has been therefore
decomposed
into each contribution $e^+e^- \rightarrow u \bar u$, $d \bar d$,
$s \bar s$, $c \bar c$ and $b \bar b$. The generated distributions
are parametrized as
\begin{equation}
D_i^{\eta} (z,M_{f0}^2) = N_i z^{\alpha_i} (1-z)^{\beta_i}
\label{her2}
\end{equation}
and analyzed using the minimization procedure MINUIT. The coefficients
$N_i$ are constrained by the normalization condition:
\begin{equation}
 \int^1_{{2 m_{\eta} \over M_{f0}}} dz \; D_i (z,M_{f0}^2) =
 {\langle n_{\eta} \rangle}_i,
\end{equation}
where the average values ${\langle n_{\eta} \rangle}_i$ are given
by HERWIG for
each quark flavor, in agreement with the total observed multiplicity
${\langle n_{\eta} \rangle}$. The parameters $N_i,{\alpha_i}$ and
${\beta}_i $ are extracted from the $\eta$ inclusive
distribution generated, for each
flavor, in the $z$ range $.025 \leq z_H \leq .95$ and shown in Table I.
As can be inferred from this table the statistical error
on the parameters is less than $5\%$.

\indent

So far we have not included the contribution from gluon.
Indeed from the analysis of the three jet events it would be possible,
in principle, to extract from HERWIG the appropriate information.
The corresponding accuracy is however unsatisfactory, due to the
limited sensitivity to hard gluon effects in $e^+e^-$ annihilation.
Then, as for the $\pi^0$ analysis,
we have followed a different approach to extract from HERWIG
the gluon fragmentation function. We have analyzed
the subprocess $ gg \rightarrow gg \rightarrow {\eta + X} $ from
$p \bar p$ annihilation at $M_{f0}=\sqrt s \sim 30$ GeV, in analogy
to the quark case. In order to
eliminate the background from the fragmentation of the spectator partons
we have constrained the etas to lye  within a cone of
semi aperture $\delta = .35 -.40$ rad
around the direction of the parent gluons emitted at $90 \deg$.
The value of $\delta$ is found by an appropriate angular study of
the generated distribution. With a parametrization of the form
{}~(\ref{her2})  we find the values of the parameters
$N_g,{\alpha_g}$ and ${\beta}_g $ given in Table II, which define Sets
I and II of fragmentation functions.

After inclusion of the gluon fragmentation function and using NLO
evolution
together with NLO terms in the $\eta$ inclusive cross
section (eqs. 3, 4) we show in Figure 1 our results at $\sqrt{S}=35$
GeV, compared with
JADE\cite{jade} and CELLO \cite{cello2} data. The agreement is satisfactory
as can be
inferred from the figure. The difference between Sets I and II is
negligeable .

After evolution to $\sqrt{S}=91.2$ GeV, we also obtain good agreement with
L3 \cite{l3} LEP data
as shown in
Fig 2.

\indent

We consider now our predictions for inclusive production in hadronic
colliders. We first compare with data from CERN ISR \cite{afs}, for
$\sqrt S=52.7$ GeV and $\sqrt S=62.4$ GeV, as shown in Figures 3 and 4
for $\mu=M=M_f=P_t$ and $\mu=M=M_f=P_t/2$ using the quark fragmentation
functions from Table I and the two gluon solutions from Table II, for
$\delta=0.35$ (Set I) and $\delta=0.40$ (Set II).
In doing so, in absence of direct data on $\eta$ production, we have
inferred  the cross section from $\pi^0$ data assuming the
experimental\cite{afs}
$\eta/\pi^0$ ratio $R$
of $0.58 \pm 0.05$ and $0.55 \pm 0.06$ respectively, indipendent from
$p_t$.
The agreement is satisfactory within
the theoretical and experimental uncertainties.

\indent

Let us focus now on the UA2 data at the $Sp\overline{p}S$ collider
\cite{ua2}. We will use two sets of quite precise $\pi^0$ data, for $P_t\leq
15$ GeV and pseudorapidity $y\simeq0$ and for $15\leq P_t \leq 45$ GeV and
$y\simeq 1.4$, and an experimental ratio $\eta/\pi^0$ of 0.5 as
obtained
from ISR data\cite{afs}. The comparison with the theoretical predictions is
shown in Figures 5 and 6 for $\mu=M=M_f=P_t/2,~P_t$ and for the two gluon
sets of fragmentation functions. The agreement is quite good at low
$p_t$, and
slightly favours set I.

On the other hand we note that at higher $p_t$ the comparison with
data  suggests a larger value for the ratio $\eta/\pi^0$, and therefore
a $p_t$ dependence for this ratio. Indeed from the result
of our previous study on inclusive $\pi^0$ production, we show in
Figure 7 the predicted $p_t$ dependence of $R$ = $\eta/\pi$ at
$\sqrt{S}$ =630 GeV, which indeed rises with $p_t$.

\indent

Finally we proceed to the predictions for LHC and SSC at
$\sqrt{S}=16$ TeV and $\sqrt{S}=40$ TeV respectively. The cross sections
are calculated at
LO (Born) and NLO and
using HMRS Set of structure functions \cite{mrs} and
are displayed in Fig. 8.\footnote{\footnotesize{The discontinuities
in the curves
are simply due to CPU time limitation on the number of the data
points.}}

To estimate the theoretical uncertainty we study in Fig. 9
the ratio of the two predictions from the two different choices
of gluon fragmentation
functions, evolved to NLO accuracy, at
$\sqrt{S}=$ 16 TeV.

We then show the theoretical ratio $\eta/\pi^0$ as
evaluated from the $\pi^0$ results of reference \cite{ftn}. As one can infer
from
Fig. 10 the ratio increases with respect to ISR energies, and shows a
dependence on $p_t$ similar to what found at $Sp\bar pS$ energies.

The uncertainty due to factorisation scheme,~especially coming from
fragmentation
functions is expected to be tiny because the
evaluation done for one jet inclusive cross section has
shown\cite{acgg} that at
collider
energies its magnitude is of the order of $5\%$ and
we can reasonably expect the same order of magnitude for one hadron inclusive
cross section. Finally, the theoretical uncertainty from the
structure functions is much smaller than that coming from fragmentation
function.

\indent

To conclude, we have performed a  complete next to leading order analysis
of inclusive
$\eta$ production in $e^+e^-$ and hadronic collisions.
We have found that the theoretical ratio $\eta/\pi^0$ depends
on $p_t$ and we have presented results for the future colliders,
where the absolute rates
can be predicted within a factor of two. This will
certainly be of help for neutral background rejection at supercolliders.
\vfill\eject
{\bf Table captions}
\vskip 24 pt

\begin{itemize}
\item{Table I:} parameters of the quark fragmentation functions as
 obtained from HERWIG in $e^+e^-$ annihilation at $M_0=30$ GeV.
\item{Table II:} parameters of the gluon fragmentation functions as
 obtained from HERWIG  at $M_0=30$ GeV, with two hypotheses on the angle
$\delta$ (see text).
\end{itemize}
%\vfill\eject
{\bf Figure Captions}
\vskip 24 pt

\begin{itemize}
\item{Fig. 1:} NLO inclusive $\eta$ production in $e^+e^-$
 annihilation with the
quark and gluon fragmentation functions evolved
at $\sqrt S= 35$ GeV, compared with data.
\item{Fig. 2:} NLO inclusive $\eta$ production in $e^+e^-$
 annihilation with the
quark and gluon fragmentation functions evolved
at $\sqrt S= 91.2$ GeV, compared with data.
\item{Figs. 3:} NLO inclusive $\eta$ production in $pp$ collisions at ISR
energies for $\mu=M=M_f= P_t,P_t/2 $ for a) $\delta =0.35$ (SetI) and
$\delta=0.40$ (Set II), see text.
\item{Figs. 4:} NLO inclusive $\eta$ production in $pp$ collisions at ISR
energies for $\mu=M=M_f= P_t,P_t/2 $ for a) Set I and b) Set II.
\item{Figs. 5:}  NLO inclusive $\eta$ production in $p \bar p$ collisions at
 $Sp \bar p S $ energies
 for $\mu=M=M_f= P_t,P_t/2 $ for Sets I and II, at
$\sqrt S=540$ GeV and $y=0$.
\item{Figs. 6:}  same as figs 5 at $\sqrt S=630$ GeV and $y=1.4$.
\item{Fig. 7:}  theoretical prediction of the ratio $\eta/\pi^0$
as function of $p_T$
at $Sp\bar pS$ energy.
\item{Fig. 8:}  NLO inclusive $\eta$ production in hadronic collisions
at LHC  and SSC energies at $y=0$.
\item{Fig. 9:}  ratio of inclusive $\eta$ cross sections for the two sets
of gluon fragmentation functions.
\item{Fig. 10:}  theoretical prediction for the ratio $\eta/\pi^0$
at LHC energy.
\end{itemize}

\vfill\eject

%%%%%%%%%%%%%%%%%%%%%%%%%%%%%%%%%%%%%%%%%%%
$$
\vbox{\offinterlineskip
\hrule
\halign{&\vrule#&
\strut\quad\hfil#\quad\cr
height4pt&\omit&&\omit&&\omit&&\omit&&\omit&\cr
& \it{Process}\hfill&&\hfill $\alpha$\hfill&&\hfill $\beta$\hfill&&\hfill
$N_q$\hfill&&\hfill$<n_{\eta}>$\hfill&\cr
height4pt&\omit&&\omit&&\omit&&\omit&&\omit&\cr
\noalign{\hrule}
height4pt&\omit&&\omit&&\omit&&\omit&&\omit&\cr
&\hfill $e^+e^-\to u\bar u$ \hfill&& \hfill $-0.91\pm 0.02$\hfill&&
 \hfill $2.09\pm 0.07$
\hfill&& \hfill $0.24$\hfill&&\hfill$0.35$\hfill& \cr
%%%%%%%%%%%%%%%%%%%%%%%%%%%%%%%%%%%%%%%%%%%%%%%%%%%%%%%%%%%%%%%%%%%%%%%%%%%
height4pt&\omit&&\omit&&\omit&&\omit&&\omit&\cr
&\hfill $e^+e^-\to d \bar d$ \hfill&& \hfill $-0.88\pm 0.02$\hfill&&
\hfill $2.14\pm 0.08$
\hfill&& \hfill $0.26$ \hfill&&\hfill$0.36$\hfill& \cr
%%%%%%%%%%%%%%%%%%%%%%%%%%%%%%%%%%%%%%%%%%%%%%%%%%%%%%%%%%%%%%%%%%%%%%%%%%%%
height4pt&\omit&&\omit&&\omit&&\omit&&\omit&\cr
&\hfill $e^+e^-\to s \bar s$ \hfill&& \hfill $-0.72\pm 0.02$ \hfill&&
\hfill $2.73\pm 0.08$
\hfill&& \hfill $0.37$\hfill&&\hfill$0.42$\hfill& \cr
%%%%%%%%%%%%%%%%%%%%%%%%%%%%%%%%%%%%%%%%%%%%%%%%%%%%%%%%%%%%%%%%%%%%%%%%%%%%
height4pt&\omit&&\omit&&\omit&&\omit&&\omit&\cr
&\hfill $e^+e^-\to c \bar c$ \hfill&& \hfill $0.14\pm 0.03$ \hfill&&
\hfill $7.10\pm 0.14$
\hfill&& \hfill $8.73$\hfill&&\hfill$0.57$\hfill& \cr
%%%%%%%%%%%%%%%%%%%%%%%%%%%%%%%%%%%%%%%%%%%%%%%%%%%%%%%%%%%%%%%%%%%%%%%%%%%%
height4pt&\omit&&\omit&&\omit&&\omit&&\omit&\cr
&\hfill $e^+e^-\to b \bar b$ \hfill&& \hfill $-0.20\pm 0.05$ \hfill&&
\hfill $11.24\pm 0.31$
\hfill&& \hfill $9.92$\hfill& &\hfill$0.69$\hfill&\cr
%%%%%%%%%%%%%%%%%%%%%%%%%%%%%%%%%%%%%%%%%%%%%%%%%%%%%%%%%%%%%%%%%%%%%%%%%%%%%
%height4pt&\omit&&\omit&&\omit&&\omit&&\omit&\cr
%&\hfill $gg \to gg$ \hfill&& \hfill $-0.37$\hfill&& \hfill $5.79$
%\hfill&& \hfill $4.55$\hfill&&\hfill$0.45$\hfill& \cr
height4pt&\omit&&\omit&&\omit&&\omit&&\omit&\cr}
\hrule}
$$
\centerline{\bf{ Table I}}\rm{
%\it{~~Table~1~Coefficients of the quark fragmentation functions~(eqs. 3-4)
%obtained through the use of
%HERWIG in $e^+e^-$ annihilation at $\sqrt{s}=30~GeV$.}
$$
\vbox{\offinterlineskip
\hrule
\halign{&\vrule#&
\strut\quad\hfil#\quad\cr
height4pt&\omit&&\omit&&\omit&&\omit&&\omit&&\omit&\cr
&\hfill ~~ \hfill&& \hfill $\delta$\hfill&&\hfill $\alpha$\hfill&&\hfill
$\beta$\hfill&&\hfill
$N_g$\hfill&&\hfill$<n_{\eta}>$\hfill&\cr
height4pt&\omit&&\omit&&\omit&&\omit&&\omit&&\omit&\cr
\noalign{\hrule}
%%%%%%%%%%%%%%%%%%%%%%%%%%%%%%%%%%%%%%%%%%%%%%%%%%%%%%%%%%%%%%%%%%%%%%%%%%%
height4pt&\omit&&\omit&&\omit&&\omit&&\omit&&\omit&\cr
&\hfill $I$\hfill&& \hfill $0.35~rad$ \hfill&& \hfill $-0.18\pm 0.06$\hfill&&
\hfill
$4.58\pm 0.25$
\hfill&& \hfill $2.52$\hfill&&\hfill$0.51$\hfill& \cr
%%%%%%%%%%%%%%%%%%%%%%%%%%%%%%%%%%%%%%%%%%%%%%%%%%%%%%%%%%%%%%%%%%%%%%%%%%%
height4pt&\omit&&\omit&&\omit&&\omit&&\omit&\cr
&\hfill$II$ \hfill&& \hfill $0.4~rad$ \hfill&& \hfill $-0.43\pm 0.06$\hfill&&
\hfill
$3.47\pm 0.26$
\hfill&& \hfill $1.48$ \hfill&&\hfill$0.62$\hfill& \cr
height4pt&\omit&&\omit&&\omit&&\omit&&\omit&&\omit&\cr}
\hrule}
$$
%\it{~~Table~2~Results of the fit to HERWIG distribution at $\sqrt{s}=630~GeV$}
\centerline{\bf{Table II}}
\vfill\eject
%%%%%%%%%%%%%%%%%%%%%%%%%%%%%%%%%%%%%%%%%%%%%%%%%

\end{document}